# Skill Evaluation for Newly Graduated Students Via Online Test

Mahdi Mohammed Younis

University of Sulaimani

Sulaimani, KRG-IRAQ

Miran Hikmat Mohammed Baban

University of Sulaimani

Sulaimani, KRG-IRAQ

*Abstract*—Every year in each university many students are graduated holding a first university degree. For example Bachelor degree in Computer Science. Most of those students have a motivation to continue with further studies to get higher education level degree in universities. However, some other students may have enthusiasm towards working based on their skills that they got during their life of studies in university. In both cases, it is required that applicant must pass a test that comprise the entire subject that they learned before. For this reasons, this research is proposing a new technique to evaluate graduate students skills.

## I. INTRODUCTION

Today's, there are lots of demand for getting a job in either government organization or public companies. Moreover, these requirements are usually asked by the people who finished a higher education degree. Also, these degrees include Diploma, Bachelor, Masters, and Doctoral. Also, It is evident that university degree makes the opportunity for getting a job much easier. However, every level needs to have one previous level. For example, it will not be possible to get matters unless they hold a bachelor degree and same for the other qualifications.

People when they get that grade, it means that they graduated from university or institution. Means completed the organizations studies successfully. After that, most of those graduated people trying to get a job or studying for a higher level of study. Furthermore, to get that opportunity most of the universities and other private organizations put a standard of restriction to make a competition between the applicants. This level of competition is testing skills through all courses that they studied during their years in universities or institutions.

Also, the most of the test that will be held in the organization and during the amount of time that will conduct to the test difficulties.

The test will comprise many subjects in the area of study and work that the applicant will apply. Thus, every graduated person will do the test in the fields they studied in their levels of educations.

On the other hand, the level of difficulties will be the same for all the testers without making different between them. However, there may be some other ways may do that, but still do the same. For example, some people may apply for networking job or some others apply for database job, in this case, there may be the various level of questions will be held.

All the test will hold on the computer in laboratory specified for the test, which provided by the specific organization. Also, all the testers must be on their computers named with their name. In the last step, the final result of the test will be shown on the screen, this means each tester will see their result with acceptance in the organization or rejected.

The question that provided are multi-choices or filling in black, but as it can notice most of these jobs in companies doing muti-choices which based on a general understanding of the working area. Furthermore, It usually starts with a simple question and level up to harder with every answer and number of question. Also, with each answer, there are points on it, for instance, the truth answer of the harder question is much more than a simple one. These points will indicate with some percentages data.

In contrast to job application process, in higher level study application process, there are lots of things to do. Moreover, one of the most important points is to take the test in many subjects. However, these subjects will be chosen by the university or the institution for instance Multimedia, Database, Networking and Artificial Intelligence.

Unlike the job application test, in this trial the form will be vary among different kinds of questions. Filling in the blank, multi-choices, writing the answers. They arranged from easier to harder. Moreover, every question has their mark, and the marks will be vary from one to another. Nevertheless, the mark will not be shown directly, and the test will hold on the paper. Later on, the exam community will check the paper exams. This one level of completion and there are lots of other comparison points to accept in the higher study.

In fact, this research will talk about how the skills can test and how to arrange the test. Also, in this propped research the following subject has mentioned:

- In the first section, a short introduction to the work and proposed idea has been written.
- General literature review and background research about the work.
- Researching some other works that have similarities with this research.
- Also, it mentioned how the work will implement and how to design the system.
- It shows how the result will show, and what the output is.

- It gives a conclusion and future work.

## II. LITERATURE REVIEW

### A. Online testing system

These days, in many universities there are lots of tastings will be held online and in different areas of study and subjects. Also, most of these tests are a multi-choices test. However, there are lots of test forms, and these forms will be vary depending on the subject and the area of study. Furthermore, some private sectors who test the applicants prefer mostly multi-choice questions. Usually, online tests are done with using computers in the laboratory that specified for the tests [1].

The online term means using internet access to do the test process, in contradiction with the idea of online. Thus, It is not mean doing the test from home, but the checking questions will through online with World Wide Web. On the other hand, the higher education test will be done on paper exam [2].

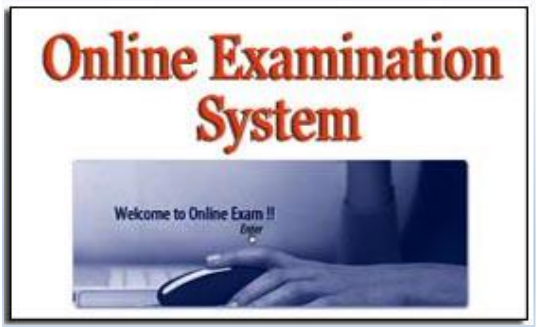

Fig. 1. Online examination web pages

### B. E- learning online system

The term of e-learning refer to electronic learning, and the most part mean learning through online methods. Such as using the internet for learning in a different level of education (schools, colleges, Institutions, and Universities). Also, there some devices these days that will put the idea on E-Learning, and those devices called Smartphones (IPad, IPod, IPhone and Galaxy et.al.) [3].

With these online study, their tests and courses available online from the much educational organization, and the mostly from universities. Moreover, those universities specify the online courses for studies called distance learning for pursuing a higher level of educational. These kinds of study will be mostly from countries who host students from other countries [4].

### C. Login System Design

This system is the most popular system among people who work or studies in different places and organizations. In fact, it is the easy way to keep member account and their secret data in safe. By using this idea, the data will not be visible to others [5]. Also, in the meaning of testing, everyone must have their page or account to conduct the test online. With doing online test the person who took the exam can see the result once finished. The next step is to close the account. From this way, no one can steal the certificate or outcome from another one [6].

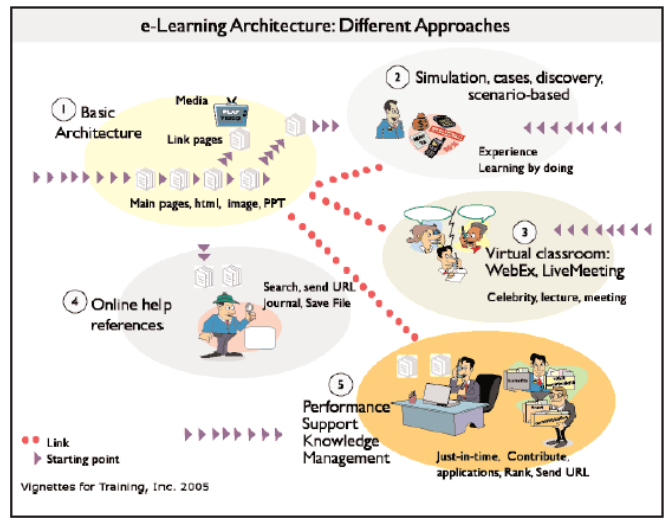

Fig. 2. E- Learning Architecture

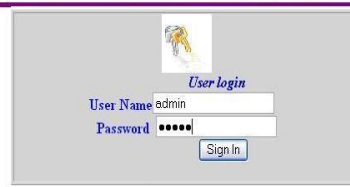

Fig. 3. Online Login System

### D. Test Storage place

The online test means doing the test online and not in the papers. This idea means there must be a place that test came from, and this place is called database storage place. From the database, the data will be loaded which means the test will be called randomly to be shown by the user end [7].

Also, when the user finished the test in specific time, all data that that related to testing such as the name of the person, answer time, Date of Birth and Address. Also, the number of correct and wrong answers will be shown as well as. Also, this storage may contain database servers or Cloud computing. Also, there are other stages such Drop Box, Sky Drive, and Google Drive [8].

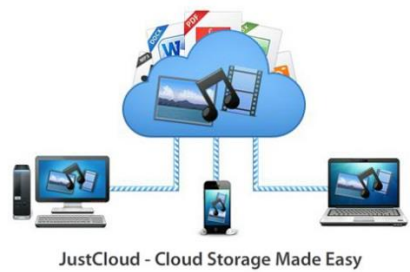

Fig. 4. Cloud Computing

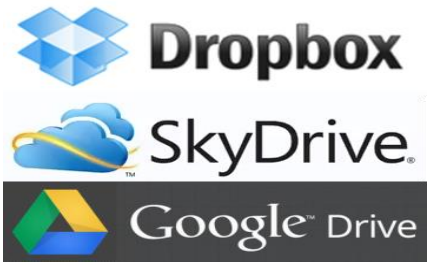

Fig. 5. Online Storage Webs

*E. Skills and job opportunities*

Nowadays, in many companies whenever a person who wants to apply for getting a job. Also, he/she will be asked to hand in lots of documents and some personal information. However, the process will not stop at this stage, but they will need to know and assess the person's skills [9]. From this way, the companies can be assured of the performance of the applicants and put in the desired position. Thus, the company will save their reputation from accepting people with lack of information and making the company going with low trends [10].

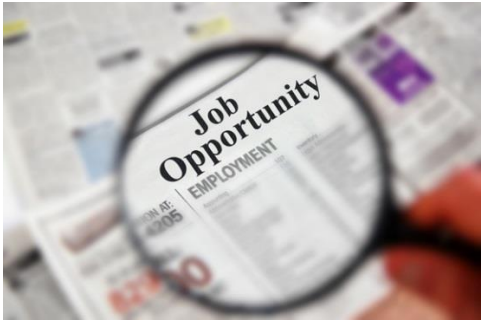

Fig. 6. Job opportunity

*F. Higher Education study*

In most of the countries, there are some obstacles for the students to study further degrees. This kind of difficulties happened due to the rules of the government of the countries. For example, in some countries the students before they accepted to study at the universities must get a high level of degree to go to the colleges. However, some colleges accept students with excellent grades such as the 90s, or 80s [11] [12].

In addition to the grade, some colleges provide obligation test to every person before acceptance in the colleges. Moreover, these test maybe on some basic subjects related to the previous skills. However, this skill test may go further hardness, especially English Academicals Qualification, academicals tests, and from this way the colleges will pick up the most suitable person who can study further [13].

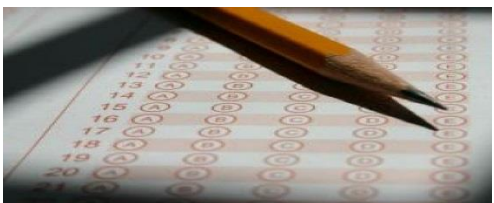

Fig. 7. Higher Education Study

*G. Exams and Time*

With every test either online or on papers, it is required to answer the question in the amount of time. Although there is time for each exam, there should be some desired calculation to each question. This idea means, there must be equality with giving time to each question [14].

However, there are differences between online test and the paper exams. Also, one of these differences is time-consuming on the test. In an online test, there is automatic decreasing clock starting from test time and begin to decrease second. Usually, the online test has less amount of time comparing to the paper tests [15].

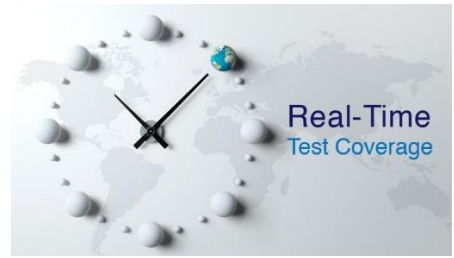

Fig. 8. Examination time and test coverage

*H. LAN Network (Local Area Network)*

This kind of networking designed for making a small communication among many different devices ranging from mobiles smartphones and IPad to computer machines and servers. Through this networking scheme, users of organization can communicate and send –receive data from each other (see figure 9).

Even through some organizations, which conducts tests either for academicals test or applicants jobs. These companies, provide a secure LAN Networking system for people who take the test, which they use login system to using the network.

Furthermore, these devices that specified for online assessments are connected to the main server that contain databases. So, they receive the question from the server, when they finished they submit the answer to the server and calculated there [16][17].

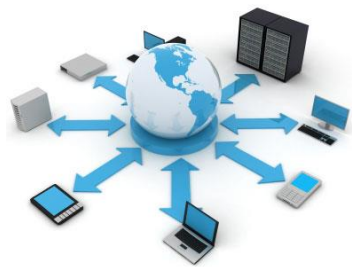

Fig. 9. LAN Networking System

*I. Database Security Issue*

It is important to provide a secure level of safety to protect database server in every organization. In some databases, there are a lot of sensitive data that related to people, either bank account, personal details or examination information tests.

In case of testing centers, those people who are working there as a staff member of the organization, they should work on protecting all the servers and databases from attackers. It happens because of some intruders may try to extract information from the server such as questions and their answers as well. This kind of fraud will leave a bad effect on the organizations and the people who do the test. In other meanings, that kind of organizations will lose their name of famousness. Moreover, people may don't get their correct grade even they do their best in the test because hackers change all the information on the server. Also, they may lose their truth opportunity to lose their job that they desire [18][19].

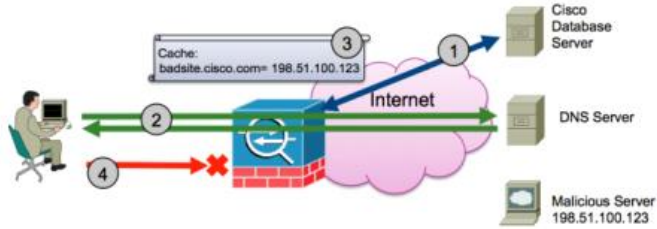

Fig. 10. Hacking Database server

*J. Online test types*

Usually, their many kinds of test can be seen on online and on papers. Questions such as multi-choices, filling in the blank, writing a brief desorption. With all of these kinds of tests, multi-choices are the most usable these days [20].

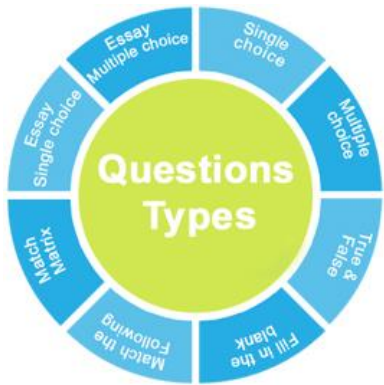

Fig. 11. Different form of questions

### III. RELATED WORK

While researching and reading some papers, it has found out some works have done in the same area, which the have little similarity with our research. However, those works used techniques different from our techniques.

*A. Related work - 1*

According to Randy Elliot, it shows how the online test work on human knowledge either it gives the advantages or drawbacks. Also, it shows the issues that face the testers during the test time.

Also, he displays the impact of online assessment in a different range of study and educational organization from elementary to university degrees.

Also, he shows that the environment of the test must be compatible in many cases. For example:

- The test should time correctly and should not exceed the number of question, which mean that the time should limit to the questions.
- The hardness and difficulties of the test should have the same range whether done online or on the paper.
- The test must not affect the sight of the testers, this means that people who do the test on computers must be able to see the texts of test clearly on small and large screen.
- The score on the test must divide equally for online and on paper.
- The most of the assessment should give enough information to the testers rather just do the normal test.

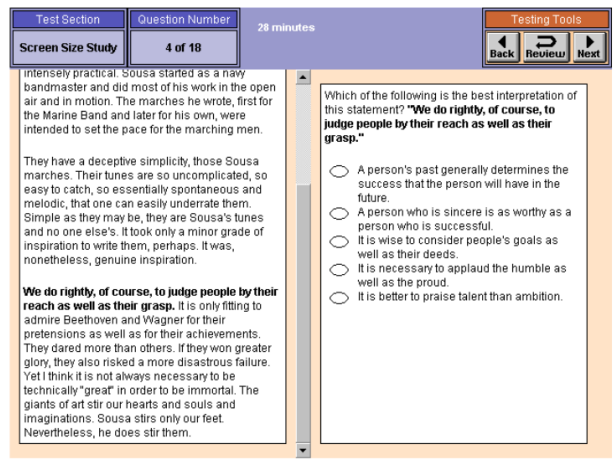

Fig. 12. Test style multi-choices

Also, it has been mentioned in his research, the most important point of the online trials:

- The correctness of the academically, which means the question must truly knowledge and not fraud data.
- The tests should be correct in grammars and spells.
- The test must have answers, and they should not vague.
- All text must be readable and not blur or hidden texts. If the texts are unclear, it will affect the user answers.

Also, in the new proposed system for the online test, it has put those point as the first important point for preparing the test for online assessment [21].

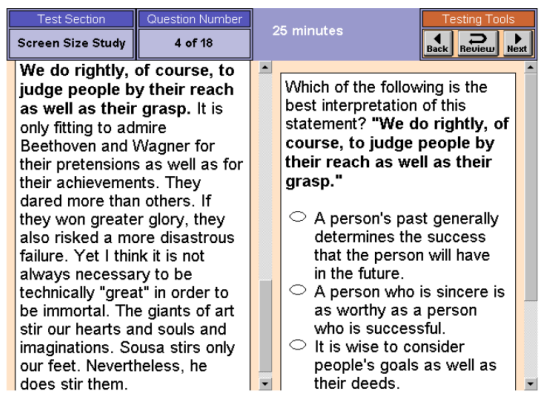

Fig. 13. Clarity of texts of the questions

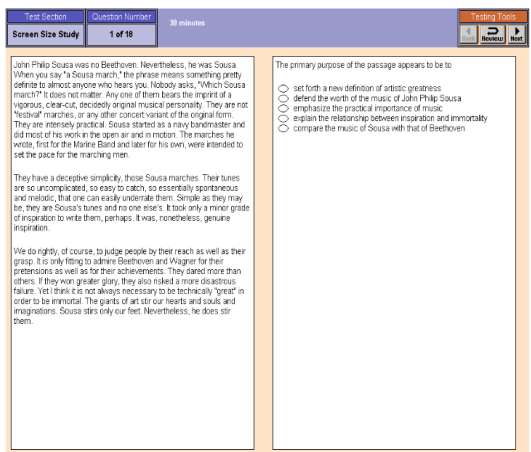

Fig. 14. Unreadable texts of the questions

*B. Related Work – 2*

According to Shachar M,& Neumann, Y., (2010). They showed that the online test had more ability to examine the capabilities of students in many different areas of studies. For example, Natural Sciences, health sciences, technologies and other studies.

- This case happens due to of the capabilities of online to test general skills in their subject area. Moreover, the question usually is multi-choices and rarely contain fill-in blanks.
- Also, those multi-choice contain four possible answers.
- Users chose only one answer between them, only one answer is possible.

Based on this research paper, it shows the effectiveness of online assessments and online materials for the students either they are graduated or still looking for further studies.

In the table below, it can be noticed that the number of people who are demanding the online sources will increase dramatically. Also, it shows the rate in percentages and starting the statistics from the year 1991 to the latest one that is 2009. Also, shows different area and level of studies. Also, in the sixth column is specified for the graduated student from universities and institutions.

Moreover, the given data encompass other levels, such as FTF Students, DE Students, Graduates, Undergraduate and another level of studies. In each level shows how many rates have got in each year as the demands for the online materials.

On the other hand, the study shows that some tests that are an available online website may name under biased test. This idea means that the test are not trusted by any academical organizations and private sectors [22].

TABLE I. TABLE DEMAND SOURCES FOR ONLINE TEST

| Period | K | ES Pos n (%) | FTF Student n(%) | DE Student n (%) | Graduate k (%) | Under Grad. + College k (%) | Other k (%) |
|---|---|---|---|---|---|---|---|
| 1991 – 1998 | 38 | 24 (63%) | 4,392 (50%) | 4,454 (50%) | 7 (18%) | 21 (55%) | 10 (26%) |
| 1999 – 2000 | 33 | 22 (67%) | 1,924 (58%) | 1,393 (42%) | 10 (30%) | 20 (61%) | 3 (9%) |
| 2001 – 2002 | 29 | 20 (69%) | 3,802 (64%) | 2,102 (36%) | 0 (0%) | 27 (93%) | 2 (7%) |
| 2003 – 2009 | 25 | 21 (84%) | 1,380 (51%) | 1,337 (49%) | 7 (28%) | 11 (44%) | 7 (28%) |
| 1991 - 2009 | 125 | 87 (70%) | 11,498 (55%) | 9,286 (45%) | 24 (19%) | 79 (63%) | 22 (18%) |

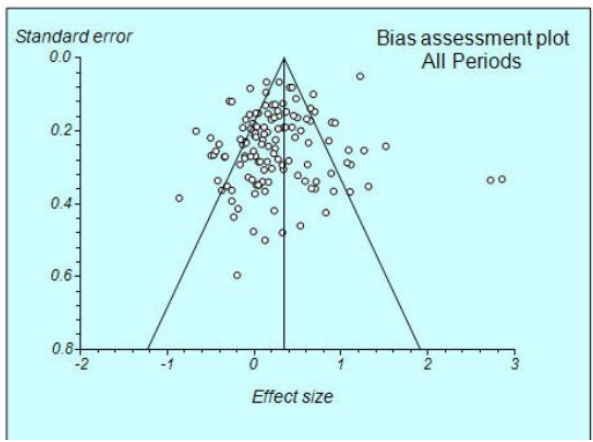

Fig. 15. The Bias with online tests

## IV. IMPLEMENTATION AND DESIGN

In our research, we have planned to propose a new systematic online test for getting applicant skills. This test consists of 50 questions in the subject area that related to the general applicant qualification. Giving examples, someone has finished bachelor degree in Computer Science; the questions will comprise general computers subjects, such as database, programming, and networking.

The form test is multi-choice with the single correct answer; this means that the tester must choose one single answer between four given answers. Thus, this kind of test mostly checks the user understanding of the subjects. However, the arrangements of the questions are not ordered according to the subjects.

The amount of the time that has been given to the test divided into the number of issues. The total time is 60 minutes; however each question must take at most 72 seconds. The graduate student should finish the test with answering all the questions in 60 minutes. After that the system will stop, there will not be any possibility to make changes to answer. Furthermore, the instructor or the lab investigator will inform the attendance to stop using a computer. Furthermore, the test will be stopped when the time reached it is final destination automatically.

At the end of the test, each person who took the test after submitted the answers they will get a page that contain the information about his/her given answers. Also, they get the final result of the test on the computer. The requirements that have been used to build this online testing system are PHP and MySQL, LAN, and wide area networking. Also, the design of the system has constructed in cooperation with the eFront learning management system.

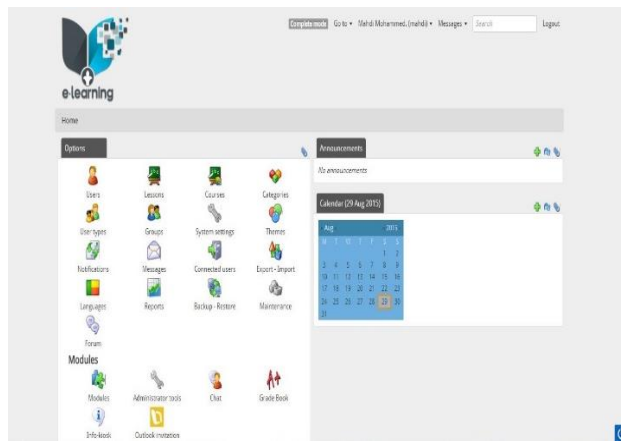

Fig. 16. System administrator panel

The same person who manage the tests and the databases who insert the questions and answers is the same employee who serves the website.

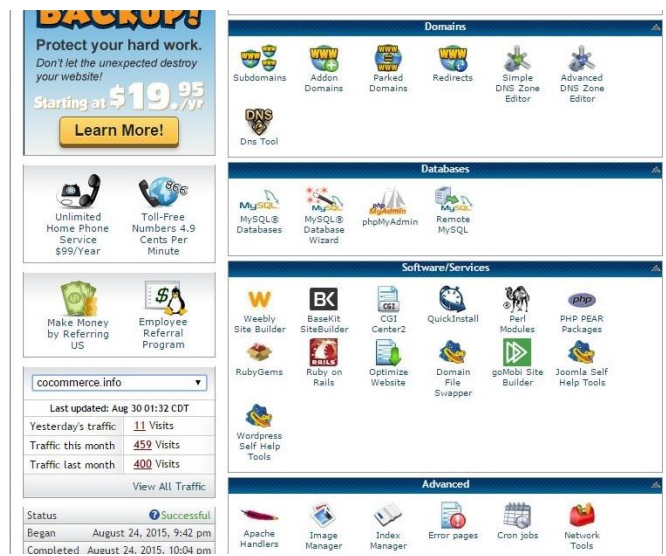

Fig. 17. Web administrator panel

In the following sections, it has been shown how the system was designed and implemented:

*A. System administrator*

The system controlled by an administrator, who control and monitor the whole system. Also, control the system database and their contents of data. Thus, all questions and answers will be entered by the admin, and the certificate will be printed out by him/her. Also, in this stage admin can add and remove users from the system.

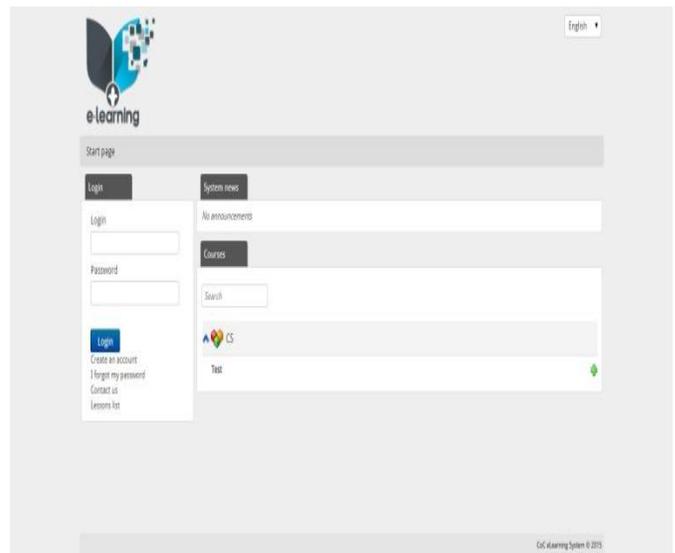

Fig. 18. Admin Login

*B. User account creation*

The graduate student will get an account to login to the system as a first stage. The reason for this security level is to protect user account from hackers. Therefore, username and password will be given them before the test.

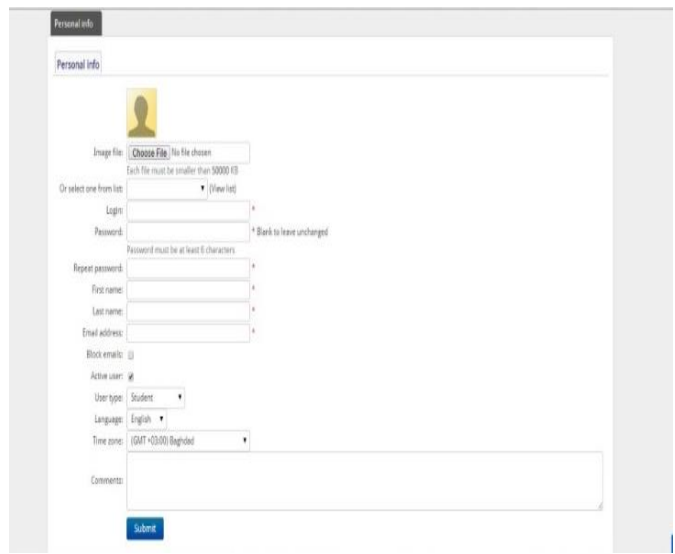

Fig. 19. User creating page

## C. Questions and answer insertion

In this stage, admin can insert the question and responses to Mysql database using insertion that specified for that matter. Also, the question can be modifying later.

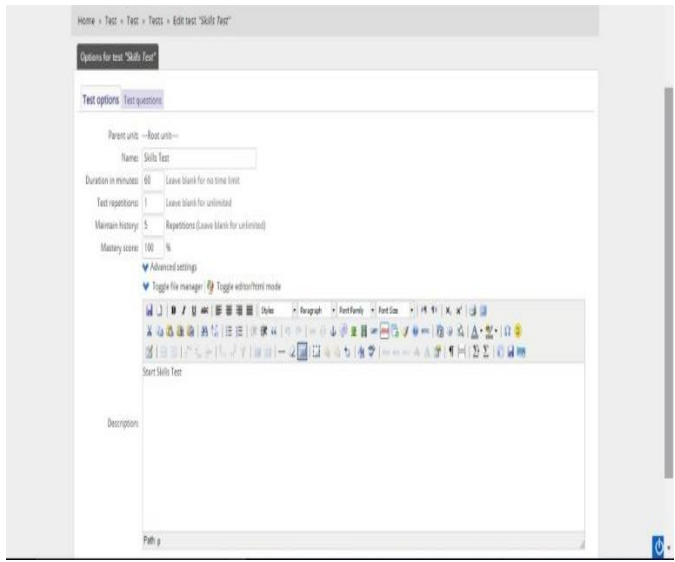

Fig. 20. Writing questions and answer

## D. User login system

On the user part, users get a web page on their computer screen; these will be asked to enter their username and password, which they got it before the test. In the case of entering wrong username or password, they will be notified with error message tells the users that they have entered wrong username or password. However, in the case of entering correct username and password they welcomed and routed them to test page with multi-choice questions.

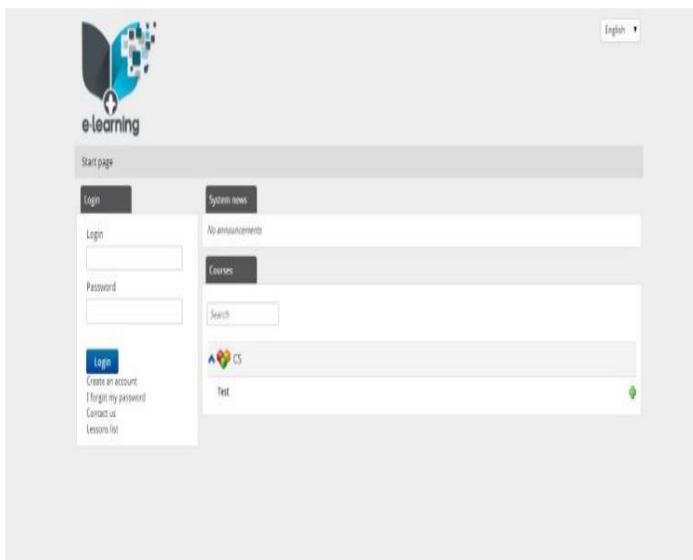

Fig. 21. User login page

## E. Test Information Page

On this page users, will get information about the test, and this information is about the time, name of the subject, start and end time and number of questions. Users that do the test is very useful to have a general idea of it; this will help them to feel more comfortable while conducting the test.

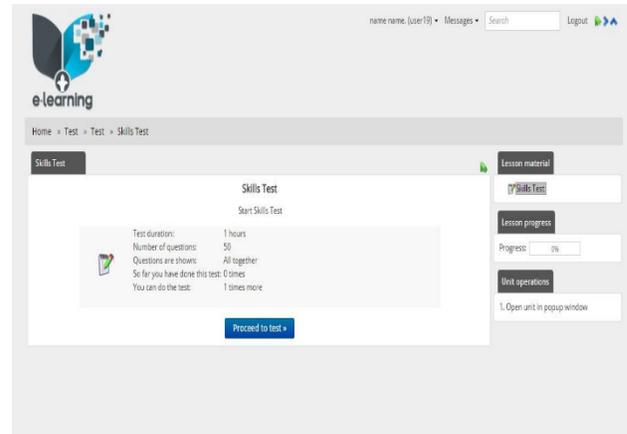

Fig. 22. Exams information page

## F. List of questions page

In this page, the registered users will get the list of multi-choice questions that consist of 50 questions from different area subjects. Also in this page it can be noticed a time specified for answering the questions, which, in fact, each question has a specific amount of time (72) seconds.

Also, the weight of score for each question is equal to 2%, and this data has been found by using the following formula:

**Score of the question * weight of score = real score**

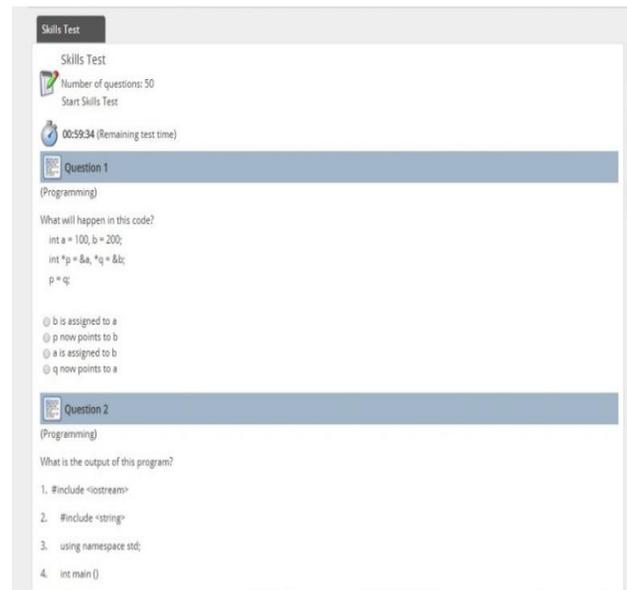

Fig. 23. First part of the question

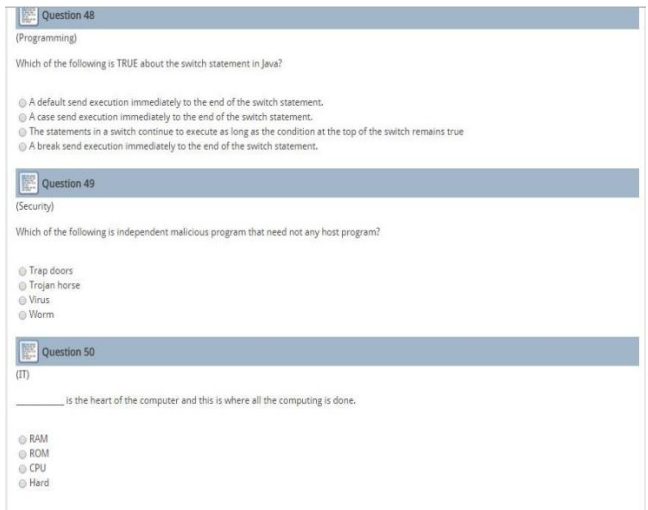

Fig. 24. Last part of the question

### G. Final Result and Feedback

At the last stage of the system, the users will get the outcome of the test at the time of submitting the answers. In this page, users can see the final score and can check their responses to know the false answers that they did in the test.

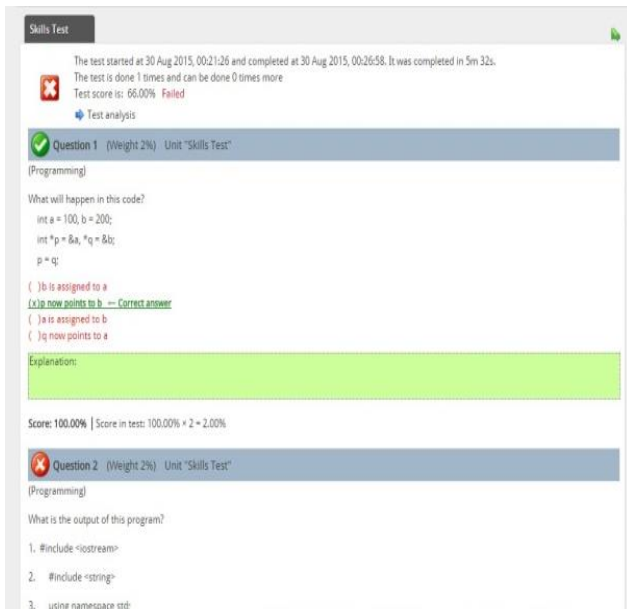

Fig. 25. Users answers feedback

### H. Online System database

All of these data, including questions, answers and user account information, are stored in a place on the server that is a database server. Also, in this database that specified for online testing systems, there are many tables. In this chart is the main place that hold all the required data about the test.

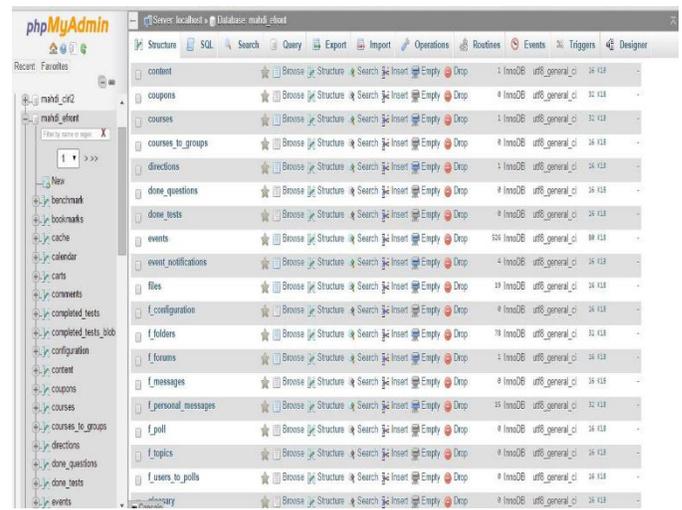

Fig. 26. Database of the system

### I. Design process flowchart

Due to the fact, students of graduating universities must do some steps to complete the system, that all these process of d The main process of our system is showing in the following flowchart:

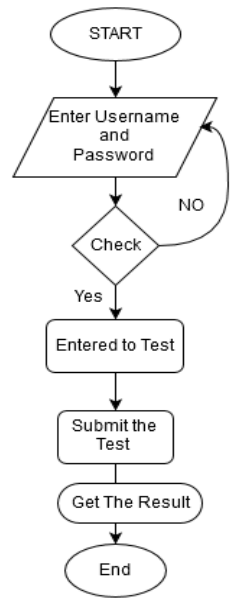

Fig. 27. Flowchart design

Users start with opening the web pages or the system. Then entering their username or password, if the username or password are correct, then they can enter the online test system successfully, if not they must re-enter the username and password correctly. After the finishing this step, user start to answer the questions and then submit the answer to the server and then they get the final result.

## V. RESULT AND DISCUSSION

As a consequence of our system, each user of the system will have their account, and can use it to track their data. Also, the user will be given a username and password before the test, to login to the test web pages. As well as the username and password, users will get a private link that is the web pages that will load from the server, and displayed on the client screen. These web pages are contained questions that considered as online tests. However, these tests are tested on both wireless and wired networks, between 10 computers. Also, we used a server as the main machine to provide the web pages and data. In fact, all these computers and even smartphones are connected to a private network. However, in our system we made the test on private and wider network. So, we had done a test for ten persons who were graduated universities in a different range of subject areas.

Each of these 10 people answered all questions, and he or she tested the system on both sides, on private and on public as well. The way that they did on the public means they went on online web pages that we put our website on that host. So, they can access the test in the lab. Also, they got their grade at the time that they submitted the test. Through the given table and statistic chart, it shows the various range of data and grades from high to low grades. Also, the charts are given, shows that how each user was doing in each particular part of the question. This shows how a particular person did on specific professional skills. For example, a user may do a test in computer science skills knowledge. Also, this test comprises lots area of computer subject such as programming, database, and networking. Also, the result and chart shows that how each person even in each section.

It is clear from the table below that their nine columns contain the name of the subjects or section for each question meaning the categorized of the question into which area is related to each question. This shows that the overall score for each person that took the test and the test score which is out of 100%. Also, the total number of the questions are 50 questions. So, from column 3 to column 7 it shows how each user did in the subcategory questions. For instance, Aram Kamal got 12 correct in programming, Networking, Database, Security and 14 IT. Also, Aram Kamal got 58.37% as an Overall grade in the test. There is the amount of time at the last column that shows when each user finished the test. In other meaning how many minutes, it took to finish the test.

If we take Aram Kamal as an example again, it finished at 58.37 minutes.

According to the following statistic bar charts (see fig 28), it is the statistical data for the online test. This test has done for the one graduated students who finished university degrees bachelor and masters in computer sciences. In this test, we have planned to evaluate the knowledge skills in different computer science subject area. We include in our test subjects, Programming, Networking, Database, Security and IT. Moreover, those ten graduate students are asked to take the evaluation test to answer 50 questions in 60 minutes.

In bar charts, we can see, different statistics values and these values divided into subject categories, and the main one is the overall score for each one. As it is clear the user named Bilal was the candidate, who got the higher score among the rest of the other students.

On the other hand, and the lowest grade that has been recorded by students named Snwr and Huner, which they got just below 50%.

Also, the person who got the average result among the other participates was named Havar, who got exactly 50%. However, the person of the test got the same result that is 70% and this grade taken by the two graduated participants named Hazhar and Haidar.

Moreover, there is another point that can which is understandable from the bar chart that is the chart for each subject. Also, everyone who has placed in the statistic bar got their rate in each area of studies. So, the subject area that recorded the highest score by the students was IT Technologies that scored about 20%. Furthermore, in the second level database, was the subject that was recorded just about to 15%.

In addition to the grade evaluations, there is another important aspect, and this aspect is the amount of time for the test and the time that used by each student. In the image, there is a different range of finishing time for the online tests. Moreover, the fastest one who finished and submitted the test was in 11 minutes and 21 seconds, this was done by a participant named Bilal. Consequently, the slowest one was Aram, who finished the test in 58 minutes and 37 seconds. Furthermore, after all of these calculation and skills evaluation we came into account that each person has an excellent skill in a range of area subjects, and lack of skills in some others.

TABLE II. ONLINE TEST RESULT

| First Name | Last Name | ProgramScore | Network Score | Data base Score | Security Score | IT Score | Final Score | Time |
|---|---|---|---|---|---|---|---|---|
| Aram | Kamal | 12 | 12 | 12 | 12 | 14 | 62 | 58.37 Sec |
| Havar | Bakhtyar | 12 | 6 | 14 | 10 | 12 | 54 | 31.49 Sec |
| Bilal | Najmaddin | 14 | 14 | 18 | 16 | 20 | 82 | 11.21 Sec |
| Haidar | Abdulrahman | 16 | 16 | 14 | 12 | 16 | 74 | 46.22 Sec |
| Hazhar | Najat | 10 | 14 | 18 | 14 | 18 | 74 | 38.11 Sec |
| Snwr | Jamal | 4 | 8 | 4 | 6 | 10 | 32 | 30.07 Sec |
| Bestan | Bahaddin | 16 | 8 | 14 | 12 | 18 | 68 | 24.03 Sec |
| Nyaz | Ali | 10 | 12 | 10 | 10 | 16 | 58 | 41.13 Sec |
| Rebwar | Rashid | 12 | 6 | 8 | 6 | 18 | 50 | 47.22 Sec |
| Huner | Hiwa | 8 | 6 | 8 | 8 | 10 | 40 | 30.39 Sec |
| Aram | Kamal | 12 | 12 | 12 | 12 | 14 | 62 | 58.37 Sec |
| Havar | Bakhtyar | 12 | 6 | 14 | 10 | 12 | 54 | 31.49 Sec |

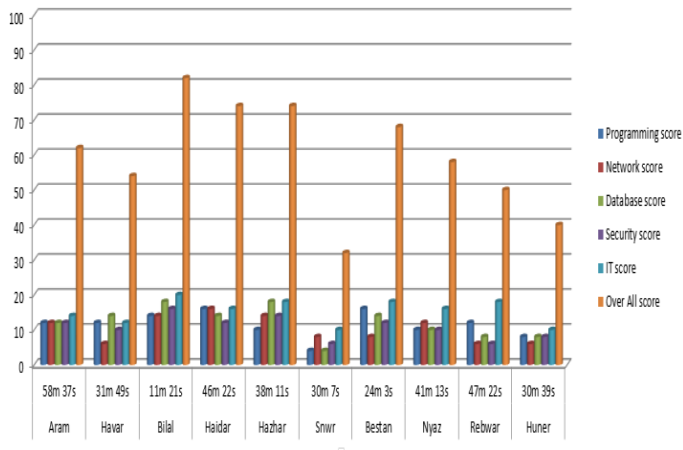

Fig. 28. User statistic results

In the table below, it shows the best skills and the worst one for each person who participated in the test. However, in some cases it can be seen that some of the test takers have more than one subject either as best skills or the worst one. Also in some other rows it says, "the rest of topics," means one of them are higher than another one. For example, the student named Aram Kamal got excellent skills in IT and the all the other area are considered as poor compare to IT subjects.

Another fact that has marked in the table is that most of the testers have got excellent skills in IT. This idea is obvious because most of the graduate students from computer sciences who want to work in company mostly adored the IT section.

In contrast to the best skill, in the poor skills column most of the applicant are suffering from lack of competencies in networking and programming as the top one, and security second one.

TABLE III. USER'S SKILLS

| Student Name | Best Skills | Poor Skills |
|---|---|---|
| Aram Kamal | IT | The rest of subjects |
| Bestan Bahaddin | IT | Networking |
| Bilal Najmaddin | IT | Programming and Networking |
| Haidar Abdulrahman | The rest of subjects | Security |
| Havar Bakhtyar | Database | Networking |
| Hazhar Najat | Database and IT | Programming |
| Huner Hiwa | IT | Networking |
| Nyaz Ali | IT | The rest of subjects |
| Rebwar Rashid | IT | Networking and Security |
| Snwr Jamal | IT | Database and Programming |
| Aram Kamal | IT | The rest of subjects |
| Bestan Bahaddin | IT | Networking |
| Bilal Najmaddin | IT | Programming and Networking |

## VI. CONCLUSION

In the final part of this research, it has concluded that there are lots of demands to get a job by the graduated student from the universities and institutions. Also, they try to get jobs according to their skills with that they got from the organization that they studied.

Also, it is quite obvious from this research that most of the companies accept their employees according to the some rules. Also, one of them is testing skills, cause they afraid of unskillful people to be a member staff of the company, by doing a specific test related to their area of study.

However, it may make difficulties for some applicants to apply, because most of them are afraid to do the test because they do not want to fail in the exam, to protect their personality. Thus, the company does not wish to accept these people with low skills.

Moreover, using test during job application will make a competent between applicants rush to get a higher score. Thus, the company by this way can choose the best one with full of knowledge in the field that they work.

Also, those companies who offer that kind of job applications, provide a security environment between their appliers and employees. So, their account information and their exam results will be safe, so it provides a good way to interest people to apply.

The university policies have the same rules for their students who apply for higher education levels. They put an exam as one of their conditions to apply. Also, this condition is the most important one because each applicant must pass the exam even they got a higher score in the other part of the rules of applying for the higher education in the universities policies.

## VII. FUTURE WORK

In the future research, there is a plan for doing this research to comprise wider area of studies:

- Getting result by SMS and E-mail
- There is a plan in the future to provide tests with a different form of testing including (filling-blank and writing description).
- Migrating to Linux server to store all the information, and to provide an area of security.
- Conducting tests on the Mobile Devices Smart Phone.

ACKNOWLEDGMENT

At the end of this research, we would like to thank University of sulaimaniyah for giving lots of facilities with computers and laboratories for doing our research. Also, we would like to send our thankful to all of our colleagues who helped us with completing this research.